\RequirePackage{ifpdf}
\documentclass[nohyper,12pt,letterpaper]{JHEP3}
\usepackage{epsfig}
\usepackage[latin1]{inputenc}
\usepackage{bbm,amsfonts}
\usepackage{graphicx}
\usepackage{amssymb,amsmath}
\usepackage{fancybox,framed}
\usepackage{dsfont}
\usepackage{mathtools}
\usepackage{braket}
\usepackage{cite}

\author{Marco S. Bianchi$^{\bf a}$
  and Matias Leoni$^{\bf b,c}$\\\\
 $^{\bf a}$ Institut f\"ur Physik,
Humboldt-Universit\"at zu Berlin,
Newtonstra{\ss}e 15, 12489 Berlin, Germany \\
  $^{\bf b}$  Physics Department, FCEyN-UBA \& IFIBA-CONICET\
Ciudad Universitaria, Pabell\'on I, 1428, Buenos Aires, Argentina \\
  $^{\bf c}$  Instituto de F\'isica de La Plata, CONICET, UNLP\
C.C. 67, 1900 La Plata, Argentina \\
  \qquad\\
  E-mail: \email{ marco.bianchi@physik.hu-berlin.de, 
  leoni@df.uba.ar}
}

\abstract{We study the three-loop four-point amplitude in ABJM theory. We determine the dual conformal invariant integrals with highest number of propagators and fix their coefficients by two-particle cuts. Evaluating such a combination of integrals in dimensional regularization we provide evidence for exponentiation of the amplitude, including the finite terms. In addition we show that the three-loop amplitude can be expressed in terms of classical polylogarithms of uniform degree of transcendentality.
}

\preprint{March 2014\\HU-EP-14/09}

\title{On the ABJM four-point amplitude at three loops and BDS exponentiation}


\csname @addtoreset\endcsname{equation}{section}

\def\bseq{\begin{subequation}}  % = 1a 1b
\def\eseq{\end{subequation}}
\def\bsea{\begin{subeqnarray}}  % = 1.1a 1.1b
\def\esea{\end{subeqnarray}}
\newcommand{\beq}{\begin{equation}}
\newcommand{\bea}{\begin{eqnarray}}
\newcommand{\eea}{\end{eqnarray}}
\newcommand{\eeq}{\end{equation}}

\renewcommand{\a}{\alpha}
\renewcommand{\b}{\beta}

\renewcommand{\d}{\delta}

\newcommand{\g}{\gamma}

\newcommand{\e}{\epsilon}

\renewcommand{\l}{\lambda}

\newcommand{\m}{\mu}

\newcommand{\n}{\nu}

\newcommand{\s}{\sigma}

\def\beq{\begin{equation}}
\def\eeq{\end{equation}}
\def\bea{\begin{eqnarray}}
\def\eea{\end{eqnarray}}
\def\Tr{\mathrm{Tr}}

\def\a{\alpha}
\def\b{\beta}
\def\g{\gamma}

\def\d{\delta}
\def\e{\epsilon}

\def\l{\lambda}

\begin{document}

\section{Introduction}
\allowdisplaybreaks

Scattering amplitudes of ABJM theory \cite{ABJM} exhibit remarkable properties and hidden symmetries, in a similar fashion as the S-matrix of ${\cal N}=4$ SYM.
This may sound rather striking since ABJM is a three-dimensional theory equipped with two gauge groups with Chern-Simons action and with bifundamental matter, which makes it quite different compared to ${\cal N}=4$ SYM.
Still, both theories are invariant under large superconformal groups, they are dual to string theory at strong coupling via the AdS/CFT correspondence and possess integrable structures, at least in the planar sector. 
Therefore the fact that some notable features of the ${\cal N}=4$ SYM S-matrix seem to have a three-dimensional counterpart in ABJM theory is not so surprising, after all.

Tree level amplitudes of ABJM can be packaged into superamplitudes of ${\cal N}=3$ superspace on-shell superfields. Such superamplitudes have been shown to enjoy Yangian \cite{BLM} and dual superconformal invariance \cite{HL2}, mirroring the similar behaviour of ${\cal N}=4$ SYM \cite{Drummond:2008aq,DHP}.
Yangian invariance calls for a possible description of the S-matrix of ABJM in terms of a Grassmannian integral formalism \cite{ArkaniHamed:2009dn,ArkaniHamed}. This was indeed shown to be the case as tree level amplitudes can be generated from an orthogonal Grassmannian integral \cite{Lee} and constructed by means of on-shell graphs \cite{Huang:2013owa}.

In ABJM theory dual superconformal invariance lacks a neat explanation at strong coupling, in contrast to ${\cal N}=4$ SYM \cite{AM,BM,Beisert:2008iq}, albeit at weak coupling it seems to be an (anomalous \cite{Bargheer:2012cp,Bianchi:2012cq}) symmetry of the S-matrix.
At loop level this symmetry is expected to manifest itself as the possibility of expressing planar amplitudes in terms of a basis of dual conformal invariant integrals  \cite{Drummond:2006rz}.
Indeed the explicit result of the four-point planar amplitude up to two loops confirms this expectation.
In particular, the cut-based construction of the amplitude \cite{CH} from a set of dual conformal invariant integrals coincides with a direct Feynman diagram computation which does not assume this property from the onset \cite{BLMPS1,BLMPS2}.

Solving the integrals underlying the computation of the amplitude reveals further interesting facts.
At one loop the four-point amplitude is subleading when evaluated in dimensional regularization \cite{ABM,CH}. The expression of the two-loop amplitude is intriguing in a number of aspects.
After proper identifications it is identical to the expression for the light-like Wilson loop with four cusps \cite{HPW,Bianchi:2013pva}. This suggests that a duality between amplitudes and Wilson loops \cite{Drummond:2007cf,Drummond:2007au,Brandhuber:2007yx} can hold in ABJM theory.
Moreover the amplitude looks strikingly similar to its one-loop counterpart of ${\cal N}=4$ SYM. 
Dual conformal symmetry and the putative duality with the Wilson loop indicate that the amplitude should satisfy anomalous conformal Ward identities, as in ${\cal N}=4$ SYM \cite{Drummond:2007au,Drummond:2007aua}. This in turn suggests that the four-point amplitude could exhibit exponentiation \cite{CH,BLMPS1,BLMPS2,Bianchi:2011aa} similarly as in the BDS ansatz \cite{Anastasiou:2003kj,BDS}.
Unfortunately no further results are available for loop amplitudes in ABJM that could support or confute the alleged duality and exponentiation. Indeed the six-point amplitude  was computed at one \cite{Bargheer:2012cp,Bianchi:2012cq,Brandhuber:2012un,Brandhuber:2012wy} and two loops \cite{CaronHuot:2012hr}, but, since it is not MHV, it would require an extension of the duality to super Wilson loops \cite{CaronHuot:2010ek}. Progress on this has been recently achieved \cite{Rosso:2014oha}.

Restricting to four points, it is fair to affirm that the duality with Wilson loops has been verified only at the lowest non-trivial order. Indeed, as recalled above, the one-loop amplitude vanishes for ABJM to finite order in dimensional regularization. Although this is in agreement with the vanishing of the corresponding Wilson loop \cite{HPW,BLMPRS}, it makes the duality rather trivial at one loop. Moreover, the fact that the one-loop amplitude is subleading does not tell us anything about its eventual role in an exponentiation ansatz, which is therefore a quite conjectural statement at this stage.
\bigskip
\bigskip

In this paper we sharpen these ideas by investigating whether exponentiation could survive at higher orders in perturbation theory and which role the one-loop amplitude plays in it.
In particular we study the four-point amplitude at three loops.
We first analyze dual conformal invariant integrals in three dimensions at three loops. Instead of constructing an explicit basis we start with the topologies with the highest number of propagators: the ladder and the tennis court.
We fix their coefficients by imposing two-particle cuts. This way we also determine integrals with a lower number of propagators, which are sensitive to this cut, and their numerators which enforce dual conformal symmetry.
Such an analysis does not fix the complete amplitude, nevertheless we conjecture that the integrals we find are sufficient to determine the amplitude or at least its maximally transcendental contributions up to finite order in dimensional regularization. This is motivated in part by analogy with lower order and ${\cal N}=4$ results and in part, a posteriori, by the remarkable properties that our ansatz exhibits, once we spell out the Mellin-Barnes representation of the relevant integrals in dimensional regularization.
First, we find that infrared divergences exponentiate to three loops. In particular the singular part is given by $1/\epsilon$ poles, coming from the product of the two-loop and the (subleading in $\epsilon$) one-loop corrections. This is consistent with the absence of contributions to the cusp anomalous dimension of ABJM at odd loop order. 
Including also the finite terms our three-loop ansatz can be written as
\begin{equation}
\mathcal{M}^{(3)}_4=\mathcal{M}^{(1)}_4\times\mathcal{M}^{(2)}_4+\mathcal{O}(\epsilon)
\end{equation}
where $\mathcal{M}^{(l)}_4$ is the 4-point $l$-loop amplitude ratio. To perform such a comparison, we expand the one-loop amplitude up to order $\epsilon^2$, since the two-loop contribution has leading $\epsilon^{-2}$ infrared poles. 
Therefore to three-loop order the logarithm of the amplitude coincides with the two-loop result up to subleading terms in $\epsilon$, such as the one-loop amplitude itself.
Of course, when exponentiating, the one-loop correction plays a crucial role in recovering the three-loop complete result.

We interpret this remarkable finding as both a signal that our ansatz is sufficient to give the whole three-loop amplitude (but not necessarily the integrand) and that the amplitude could exponentiate. 
Indeed it is tempting to conjecture that this pattern could hold beyond three-loop order and that the whole perturbative series exponentiates.
The fact that this is the case at three loops can be seen as an indirect test of dual conformal invariance and duality with Wilson loops, since exponentiation can be interpreted as a consequence of a (dual) conformal Ward identity. 

The available two-loop results for amplitudes in ABJM exhibit uniform transcendentality. This is true even for non-planar corrections \cite{Bianchi:2013iha,Bianchi:2013pfa} and seems to be a property of the ABJM S-matrix, at least at four points. Recently this intuition has been given an explanation in terms of a construction of amplitudes using on-shell graphs \cite{Huang:2014xza} similar to that for ${\cal N}=4$ SYM \cite{ArkaniHamed:2012nw}.
Accordingly, we expect the three-loop amplitude to be expressed in terms of transcendental functions of uniform degree three.
Indeed, by exploiting the identification with the higher order expansion of the one-loop amplitude, we are able to provide an explicit form for our ansatz at three loops (and also for some symmetrized combinations of ladder and tennis court integrals, respectively). Such an expression is indeed in terms of classical polylogarithms with uniform degree of transcendentality. Their arguments are in terms of square roots of ratios of kinematic invariants.

\section{Dual conformal invariant integrals in three dimensions}\label{sec:1}

Tree level superamplitudes in the ABJM model are Yangian \cite{BLM,DHP} and dual superconformal \cite{HL2,Drummond:2008aq,Brandhuber:2008pf} invariant. This symmetry can be checked explicitly for lower multiplicity amplitudes and proved to extend to arbitrary number of points via recursion relations \cite{GHKLL}, generalizing the BCFW construction \cite{BCF,BCFW} to ABJM.
In this paper we only focus on the four-point superamplitude. Its tree level expression reads
\begin{equation}
{\cal A}^{(0)}_4 (\bar{1}, 2, \bar{3}, 4) = i\, \frac{\delta^{(6)}(Q)\delta^{(3)}(P)}{\braket{1 2}\braket{2 3}}
\end{equation}
where $P$ and $Q$ are the total momentum and supercharge and the $\delta$ functions enforce momentum conservation and supersymmetry.
We follow the conventions of \cite{Brandhuber:2012un}, which are reviewed in the appendix.
Through unitarity dual conformal invariance propagates to loop level corrections ${\cal M}_4^{(l)}$ in the planar limit, which will be assumed throughout this paper
\begin{equation}
{\cal A}_4 = {\cal A}^{(0)}_4\, \left(1 + \sum_{l=1}^{\infty}\, {\cal M}_4^{(l)} \right) = {\cal A}^{(0)}_4\, \left(1 + \sum_{l=1}^{\infty}\, \lambda^l\, M_4^{(l)} \right) 
\end{equation}
where $l$ stands for the loop order, $\lambda \equiv \frac{N}{4\pi\,k}$ is the 't Hooft coupling and ${\cal M}$ denotes ratios between loop corrections and the tree level superamplitude.
 
For ABJM theory dual conformal invariance is still lacking a precise strong coupling interpretation \cite{Adam:2009kt,Adam:2010hh,DO,Bakhmatov:2010fp,Bakhmatov:2011aa,OColgain:2012si} in terms of a self duality of the dual sigma model under fermionic T-duality, as occurs for ${\cal N}=4$ SYM \cite{AM,BM,Beisert:2008iq}. Nevertheless explicit analysis of amplitudes at weak coupling suggests that it is a (eventually broken by infrared divergences) symmetry of the on-shell sector of ABJM in the planar limit.
As a consequence we expect all planar loop corrections to be expressible in terms of a basis of dual conformal invariant integrals. Once such a basis is known, the coefficients can be fixed by (generalized) unitarity \cite{BDDK,BDK,BCF2}, simplifying the determination of amplitudes extremely. This motivates the importance of classifying dual conformal integrals in three dimensions. 

Searching for such integrals is not as straightforward as in four dimensions, especially at odd loops, where the number of integrations is odd \cite{HL2}.
An elegant way to describe them exploits a five-dimensional formalism \cite{CH}. In this setting dual coordinates describing momenta are embedded into five-dimensional variables $X$. External points are taken on the light-cone of five-dimensional Minkowski space, and loop variables are projectively reduced to three-dimensional integrals. Within this setting dual conformal invariance translates into Lorentz symmetry and scale invariance of the integrand with respect to each point $X_i$, both external and internal.
At one loop this criterion uniquely determines the integral (in five-dimensional formalism)
\begin{equation}
I^{(1)} = \int {\cal D} X_5\, \frac{\varepsilon\left( X_5, X_1, X_2, X_3, X_4 \right)}{X_{51}^2\, X_{52}^2\, X_{53}^2\, X_{54}^2}
\end{equation}
where we use the notation $\varepsilon \left( \dots \right)$, understanding contraction of Lorentz indices and ${\cal D} X_5$ for the projective integration measure \cite{CH}.
This is the integral appearing in the one-loop amplitude, as verified by a direct unitarity-based computation \cite{CH}
\begin{equation}
M^{(1)}_4\, = \, i\, I^{(1)}
\end{equation}
After projectively reducing it to three dimensions (as spelled out in \cite{CH}) and going to momentum space it reads
\begin{equation}\label{eq:boxfunction}
I^{(1)} \equiv \int \frac{d^dl}{(2\pi)^d}\, \frac{N(l)}{l^2\, (l-p_1)^2\, (l-p_{12})^2\, (l+p_4)^2}
\end{equation}
where $p_{12} \equiv p_1+p_2$.
Since it will appear several times in this paper, we define the numerator as
\begin{equation}\label{eq:num}
N(l) \equiv s\, \Tr \left(l\, p_1\, p_4 \right) + l^2\, \Tr \left( p_1\, p_2\, p_4 \right)
\end{equation}
where the trace $\Tr$ is over spinor indices and, e.g., $\Tr(p_1\, p_2\, p_4) = 2\, \varepsilon \left(p_1,p_2,p_4 \right)$.
It is more convenient to think at this numerator as the three-dimensional reduction of the unique five-dimensional one. Since the reduction can be performed in four equivalent ways focusing on the loop momentum flowing in each of the edges of the box, we have a set of alternative manners of expressing it in three dimensions. 
These choices correspond to identities like
\begin{equation}\label{eq:identitynum}
N(l) \equiv s\, \Tr \left(l,p_1,p_4\right) + l^2\, \Tr \left(p_1,p_2,p_4\right) =
t\, \Tr \left(l,p_1,p_2\right) + (l-p_1)^2\, \Tr \left(p_1,p_2,p_4\right)
\end{equation}
which can be derived from spinor algebra.
In practice we use this freedom to select a form of the numerator which fits the computation best.

The one-loop amplitude shows the ubiquitous presence of $\varepsilon$ tensors appearing in dual conformal invariant integrals in three dimensions.
In the integral \eqref{eq:boxfunction} we have set generic dimension $d$, since we shall eventually perform its computation within dimensional regularization $d=3-2\epsilon$. This point is extensively discussed in section \ref{sec:regularization}. Within this setting the one-loop box integral \eqref{eq:boxfunction} is subleading in the dimensional regularization parameter and so is the one-loop amplitude.

At two loops there are four independent dual conformal invariant integrals for four-particle scattering processes \cite{CH}, whose topologies are depicted in figure \ref{fig:DCI2}.
\FIGURE{
\centering
\includegraphics[width=1.0\textwidth]{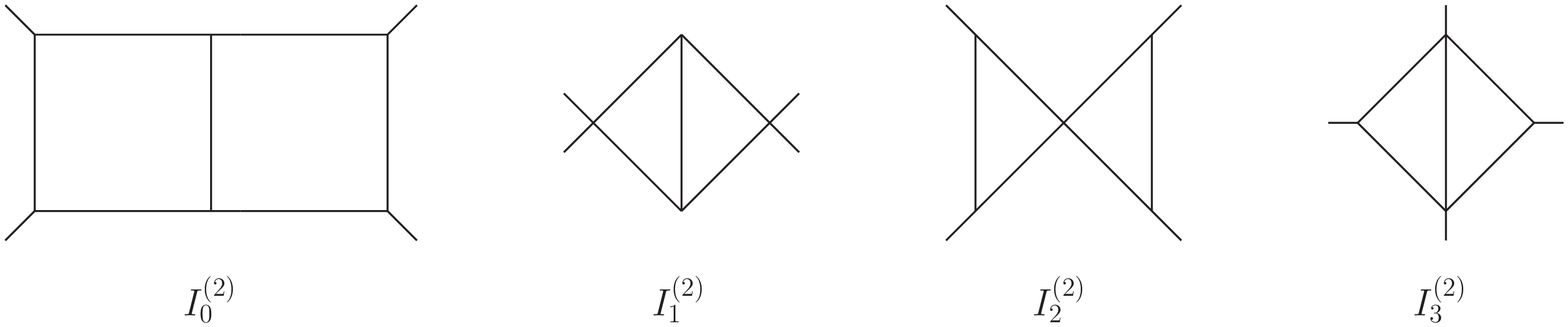}
\caption{Two-loop dual conformal invariant integrals for four-point amplitudes.}
\label{fig:DCI2}
}
Though not necessary, these integrals can be also identified in five-dimensional formalism. In this setting it is natural to express the numerator of the double-box in terms of $\varepsilon$ tensors.  
Explicitly it reads
\begin{equation}\label{eq:doublebox}
I^{(2)}_{0,s} \equiv \int\frac{d^dl}{(2\pi)^d}\frac{d^dk}{(2\pi)^d}\, \frac{N(l_1)\, N(l_2)}{t\,l_1^2 (l_1-p_{12})^2 (l_1-p_1)^2 l_2^2 (l_2+p_4)^2 (l_2-p_{12})^2 (l_1-l_2)^2}
\end{equation}
where again we have used the numerators \eqref{eq:num}.
Of course the pairs of $\varepsilon$ tensors contained in the numerator of \eqref{eq:doublebox} can be reduced to scalar products, leading to more familiar numerators \cite{CH}.
Nevertheless the expression \eqref{eq:doublebox} appears simpler and more natural from the point of view of unitarity.
Indeed one can reconstruct the two-loop amplitude starting by considering its two-particle cuts, separating it into a tree and a one-loop four-point amplitudes. Such a construction has been explicitly performed in \cite{Bianchi:2013pfa}. Using a Schouten identity such a cut immediately coincides with that of the double-box \eqref{eq:doublebox}, up to an additional piece which is recognized as coming from the double-triangle $I^{(2)}_1$ of figure \ref{fig:DCI2}, which reads
\begin{equation}\label{eq:doubletriangle}
I^{(2)}_{1,s} \equiv \int\frac{d^dl_1}{(2\pi)^d}\frac{d^dl_2}{(2\pi)^d}\, \frac{s^2}{l_1^2 (l_1-p_{12})^2 l_2^2 (l_2-p_{12})^2 (l_1-l_2)^2}
\end{equation}
This cut analysis already excludes the integral $I^{(2)}_2$ of figure \ref{fig:DCI2}, which is sensitive to the two-particle cut and would have been detected. The coefficient of the last integral $I^{(2)}_3$ cannot be fixed at this stage. One can then consider a three-particle cut. It has to vanish since it separates the two-loop amplitude into two five-point tree level ones, which are zero in ABJM. This condition entails that $I^{(2)}_3$ does not contribute to the amplitude.
Henceforth the two-loop expression of the four-point amplitude in terms of integrals reads
\begin{equation}\label{eq:2loopint}
M^{(2)}_4 = I^{(2)}_{0,s} + I^{(2)}_{0,t} + I^{(2)}_{1,s} + I^{(2)}_{1,t}
\end{equation}
where integrals in the $t$-channel can be read from \eqref{eq:doublebox} and \eqref{eq:doubletriangle} after a cyclic permutation of external momenta labels by one site.
By computing the double-box and double-triangle integrals in dimensional regularization one obtains the two-loop amplitude
\begin{equation}\label{eq:2loopamp}
M^{(2)}_4 = -\frac{(-s/\m'^2)^{-2 \epsilon }+(-t/\m'^2)^{-2 \epsilon }}{(2 \epsilon)^2} + \frac{1}{2} \log ^2 \frac{s}{t} + \frac{2 \pi ^2}{3} + 3 \log^2 2 + {\cal O}(\epsilon)
\end{equation}
where the dimensional regularization mass scale has been redefined as $\mu'^{2}= 8 \pi e^{-\gamma_E}\mu^2$.
This expression strikingly resembles the one-loop amplitude in ${\cal N}=4$ SYM, up to straightforward identifications \cite{CH,BLMPS1,Bianchi:2011aa}.
As for ${\cal N}=4$ SYM \cite{Drummond:2007aua,Drummond:2007cf,Drummond:2007au,Brandhuber:2007yx}, it happens to coincide with the expectation value of a light-like Wilson loop with four cusps \cite{HPW,Bianchi:2013pva,BLMPRS}. This hints at a possible amplitude/Wilson loop duality for ABJM. For the bosonic Wilson loop such a relation would hold for four-point amplitudes only. Indeed, by analogy with ${\cal N}=4$ SYM we expect such a duality to be valid for MHV amplitudes only, and in ABJM only four-point amplitudes can be regarded as MHV in terms of the Grassmann variables of ${\cal N}=3$ superspace \cite{BLM}. A possible extension to higher point amplitudes would require considering super Wilson loops \cite{CaronHuot:2010ek}. Progress in such a construction has been recently carried out in \cite{Rosso:2014oha}.

It has to be noticed that the two-loop amplitude \eqref{eq:2loopamp} is mostly given by the double-box, which contains the leading infrared poles and the functions with highest degree of transcendentality.
Despite this integral fails to be uniformly transcendental by itself, all its lower transcendentality terms are exactly cancelled by the double-triangle.
In addition the double-box and double-triangle integrals on their own suffer from unphysical off-shell infrared divergences. This is the integrals are already divergent off-shell, as can be seen at the integrand level because of the presence of internal three-point vertices.
Since we expect infrared divergences of the amplitude to be produced by soft and collinear regions of the integration due to massless external particles, we want the aforementioned singularities to drop off from the result.
This is precisely the case, as spurious off-shell divergences eventually cancel between the double-box and the double-triangle \cite{BLMPS2}.
This suggests that their combination is somehow the most natural integral to be considered. The fact that it possesses uniform transcendentality would simplify its solution considerably, if one could express it in terms of uniformly transcendental master integrals (as in \cite{Brandhuber:2013gda} for form factors), as occurs successfully in four dimensions \cite{Henn:2013pwa}.

\section{Three-loop dual conformal invariant integrals}

At three loops determining all dual conformal invariant integrals, is rather cumbersome, even in the five-dimensional formalism.
Moreover in some cases there are several inequivalent numerators enforcing dual conformal invariance, for the same topology.

We shall not perform here a complete classification of three-loop dual conformal invariant integrals. Rather, we inspect some necessary condition they have to satisfy, and use them to construct the dual conformal invariant integrals possessing the maximal number of propagators. These are constructed with trivalent vertices only and have $3L+1$ propagators, which is ten at three loops. Guided by analogy with the two-loop computation, it is possible that their maximally transcendental part could suffice to determine the whole three-loop amplitude and that all other integrals just contribute to restore uniform transcendentality and cancel unphysical divergences.

As recalled in the previous section, we can use a five-dimensional formalism where dual conformal invariance translates into Lorentz symmetry and scale invariance with respect to any dual space variable $X_i$.
On the one hand at three loops there are nine integrations. On the other hand the number of inverse powers of $X$ from the propagators is even. Hence we have to require an odd number of $X$'s in the numerator in order to ensure scale invariance. This can be accomplished in a Lorentz invariant manner by contractions with an odd number of $\varepsilon$ tensors. 
The amount of such tensors is limited by how many powers of loop variables are needed in the numerator in order to guarantee scale invariance. In turn, this depends on the kind of sub-integral the given dual coordinate $X_l$ is involved in. Bubble sub-integrals have only two powers of the loop variable in the denominator but three in the numerator due to the integration measure. They can not give scale invariance and are therefore excluded on the grounds of dual conformal symmetry. Triangle sub-integrals do not require any additional powers of their internal coordinate in the numerator, box sub-integrals demand one power and pentagons two.
At three loops and with four external points these are the only sub-integrals we can construct.
In particular the topologies with trivalent vertices only are the ladder and the tennis court integrals, depicted in figure \ref{fig:lad-tc}.
\FIGURE{
\centering
\includegraphics[width=0.8\textwidth]{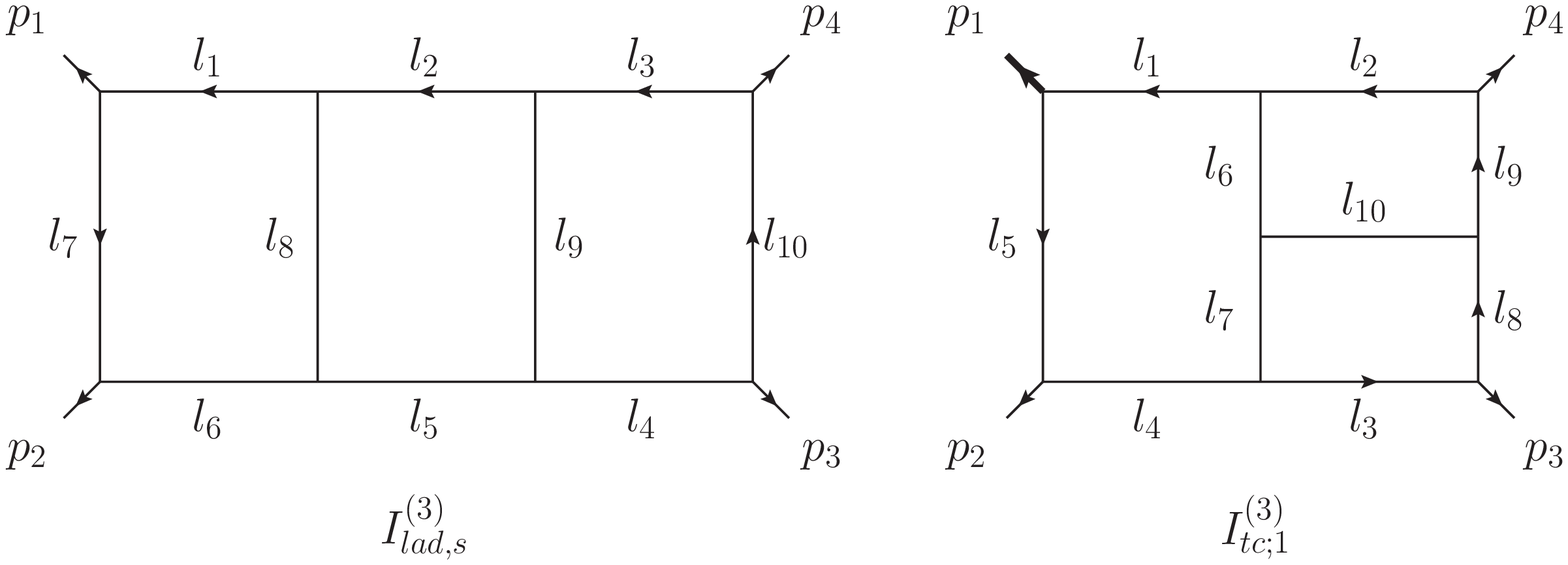}
\caption{Ladder and tennis court integrals.}
\label{fig:lad-tc}
}
The former requires one power of each internal coordinate in the numerator. The latter needs one additional power of the dual variable corresponding to the pentagon sub-integral.
In both cases we can have either three or one $\varepsilon$ tensors in the numerator, depending on how contractions are performed. 
As anticipated, there in not a unique choice for such numerators. 
We show here one particular election, guided by intuition, which turns out to be convenient for unitarity-based computation. Going to the more familiar representation in three-dimensional momentum space our ladder integral reads
\begin{equation}\label{eq:ladders}
I^{(3)}_{lad,s} \equiv \int \frac{d^d l_1 d^d l_2 d^d l_3}{(2\pi)^{3d}}\, \frac{N(l_1)\, N(l_2)\, N(l_3)}{t^2\, l^2_1 l^2_2 l^2_3 l^2_4 l^2_5 l^2_6 l^2_7 l^2_8 l^2_9 l^2_{10}}
\end{equation}
where momenta correspond to those of figure \ref{fig:lad-tc}.
Our tennis court is given by
\begin{equation}\label{eq:tenniscourt}
I^{(3)}_{tc,1} \equiv \int \frac{d^d l_2 d^d l_4 d^d l_5}{(2\pi)^{3d}}\, \frac{N(l_1)\, N(l_2)\, N(l_3+p_{12})\, (l_1+p_4)^2}{s\, t^2\, l^2_1 l^2_2 l^2_3 l^2_4 l^2_5 l^2_6 l^2_7 l^2_8 l^2_9 l^2_{10}}
\end{equation}
where the numerators $N$ were defined in \eqref{eq:num} and we refer to the discussion below that formula for explanations.
The label indicates the position of external momenta as in figure \ref{fig:lad-tc}, where the index corresponds to the thick line.

We observe that the ladder integral we have defined exhibits a suggestive pattern when compared to the box and double-box ones.
This calls for proposing an $L$-loop dual conformal invariant three-dimensional ladder
\begin{equation}\label{eq:Lladder}
\int \prod_{i=1}^L\,\left( \frac{d^{d}\, l_i}{(2\pi)^{d}}\, \frac{1}{l_i^2\, (l_i-p_{12})^2\, }\right)\, \prod_{j=1}^{L-1}\, \frac{1}{(l_j-l_{j+1})^2}\, \frac{1}{(l_1-p_1)^2\, (l_L+p_4)^2}
\end{equation}
We shall comment more on this multi-loop ladder integral in the next section.

Unlike ${\cal N}=4$ SYM, there are several topologies with a lower number of propagators, which support numerators enforcing dual conformal invariance and which can also contribute to the amplitude.
Those which are sensitive to two-particle cuts will be automatically identified performing such a cut on the ladder and tennis court and imposing consistency with the same cut on the amplitude.
With such an analysis, that we shall carry out explicitly in the next section, we find the following integrals, which are depicted in figure \ref{fig:desc}.
\FIGURE{
\centering
\includegraphics[width=1.\textwidth]{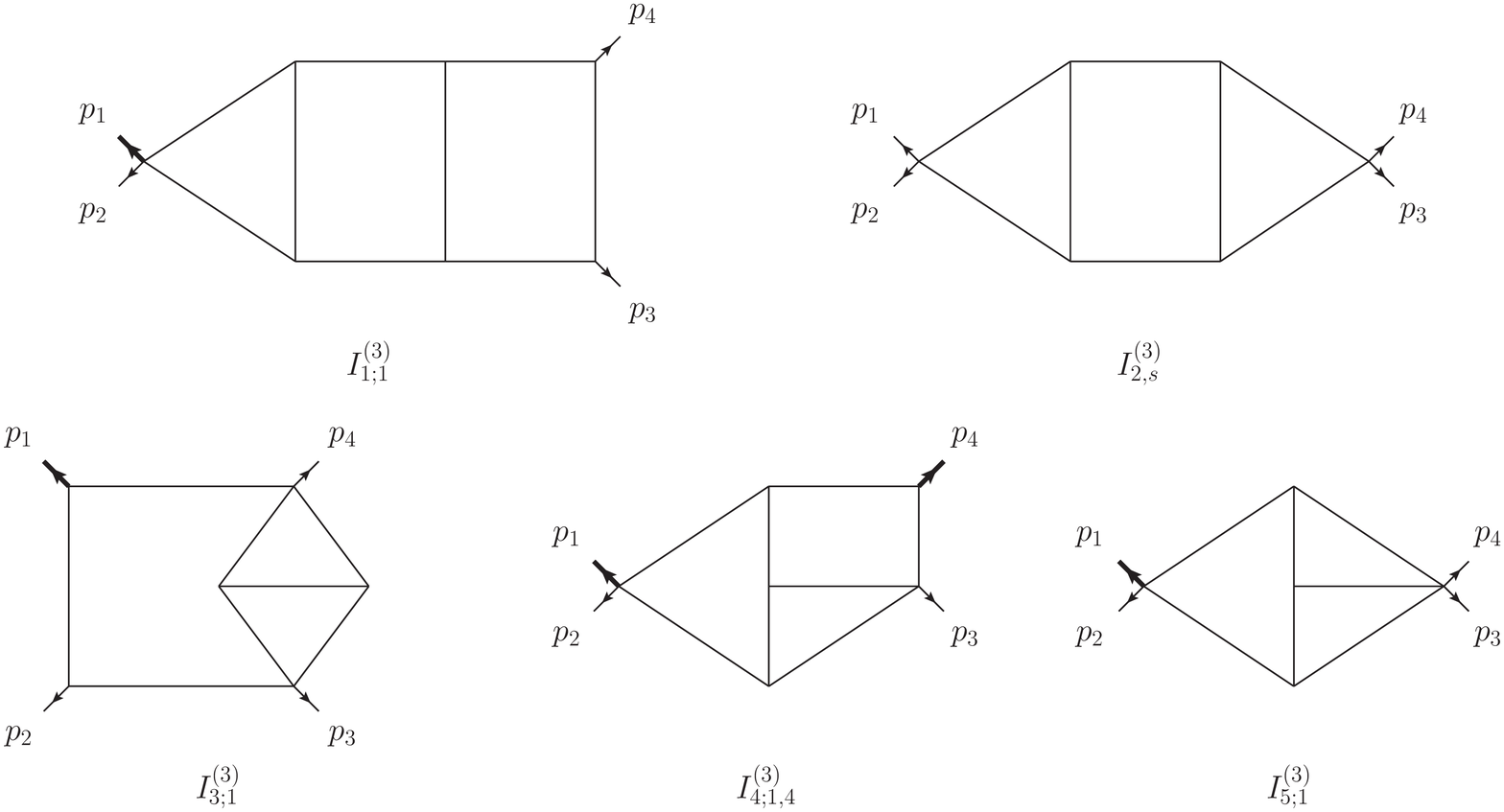}
\caption{Some dual conformal invariant integrals coming from eliminating propagators in the ladder ($I^{(3)}_{1}$ and $I^{(3)}_{2}$) and the tennis court ($I^{(3)}_{3}$, $I^{(3)}_{4}$ and $I^{(3)}_{5}$).}
\label{fig:desc}
}
The first two emerge contracting some edges of the ladder topology and their expressions read (with momenta labelled as for the ladder integral \eqref{eq:ladders})
\begin{align}
I^{(3)}_{1;1} &\equiv\, \int \frac{d^d l_1\,d^d l_2\,d^d l_3}{(2\pi)^{3d}}\, \frac{s^2 \, N(l_2)\, (l_1+p_4)^2}{t\, l^2_1 l^2_2 l^2_3 l^2_4 l^2_5 l^2_6 l^2_8 l^2_9  l^2_{10}}
\nonumber\\
I^{(3)}_{2,s} &\equiv\, \int \frac{d^d l_1\,d^d l_2\,d^d l_3}{(2\pi)^{3d}}\, \frac{s^2\, N(l_2)}{t\, l^2_1 l^2_2 l^2_3 l^2_4 l^2_5 l^2_6 l^2_8 l^2_9}
\end{align}
The label in the $I^{(3)}_{1;a}$ integral stands for the first external momentum $p_{i,i+1}$ flowing out of the four-fold vertex. 
The remaining ones can be seen to descend from the tennis court and they are  (with momenta labelled as in \eqref{eq:tenniscourt})
\begin{align}
I^{(3)}_{3;1} &\equiv\, \int \frac{d^d l_1\,d^d l_2\,d^d l_3}{(2\pi)^{3d}}\, \frac{N(l_1)\, (l_1 + p_4)^2}{l^2_1 l^2_4 l^2_5 l^2_6 l^2_7 l^2_8 l^2_9 l^2_{10}}
\nonumber\\
I^{(3)}_{4;1,4}&\equiv\, \int \frac{d^d l_1\,d^d l_2\,d^d l_3}{(2\pi)^{3d}}\, \frac{s\, N(l_2)\,  (l_1+p_4)^2}{t\, l^2_1 l^2_2 l^2_3 l^2_4 l^2_6 l^2_7 l^2_9 l^2_{10}}
\nonumber\\
I^{(3)}_{5;1}&\equiv\, \int \frac{d^d l_1\,d^d l_2\,d^d l_3}{(2\pi)^{3d}}\, \frac{s\, N(l_1)}{t\, l^2_1 l^2_2 l^2_3 l^2_4 l^2_6 l^2_7 l^2_{10}}
\end{align}
Again extra labels are used to fix the position of external legs in the integral, similarly to the previous ones. In $I^{(3)}_{4;i,j}$, $j$ is the momentum flowing into the box sub-integral and can only take values $i+2$ or $i-1$, producing two inequivalent integrals.

We stress that these are not all the dual conformal invariant three-loop integrals in three dimensions.
In particular the inspection of two-particle cuts overlooks those integrals which do not possess this kind of cuts. These are identified by requiring that there are no consecutive three-point vertices with an external line in their topology.
This class of integrals is a restricted subset of dual conformal invariant ones and one could attempt to classify them all, however we shall not carry this out here.

\section{Two-particle cuts}

As discussed in the above section, the integrals with the largest number of propagators we expect to contribute are the ladder and the tennis court. 
\FIGURE[l]{
\centering
\includegraphics[width=0.4\textwidth]{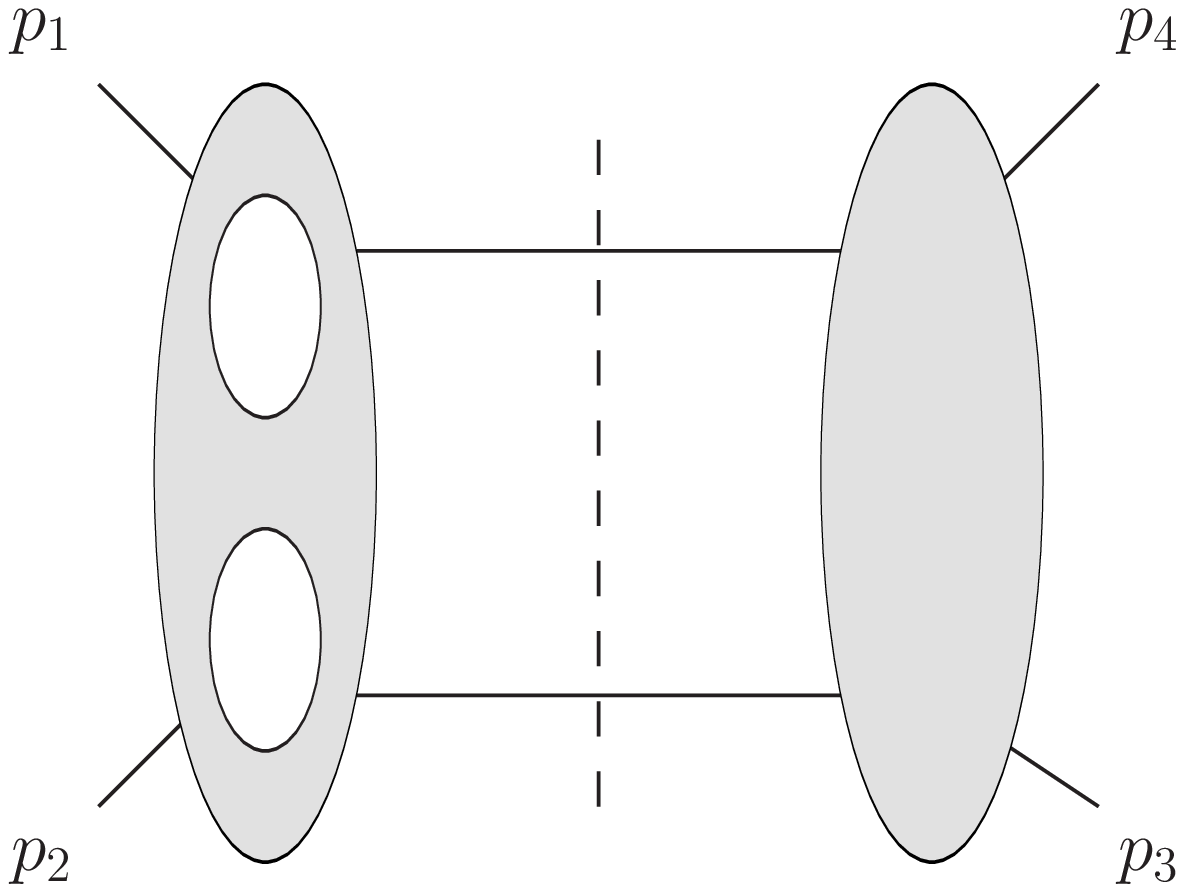}
\caption{Two-particle cut, $s$-channel.
}
\label{fig:cut}
}
They can be both identified with a two-particle cut, as in figure \ref{fig:cut} for the $s$-channel. This separates the three-loop amplitude into a tree level and a two-loop four-point amplitudes.
We shall refer to the cut momenta as $k$ and $k-p_{12}$, satisfying the cut conditions $k^2=0$ and $(k-p_{12})^2$. Hence the four-point sub-amplitudes depend on the two invariants $s$ and $(k-p_1)^2$.
Considering the integral topologies, we expect the part of the two-loop amplitude \eqref{eq:2loopint} containing integrals $I_{0,s}^{(2)}$ and $I_{1,s}^{(2)}$ to give rise to the ladder topology, whereas the $I_{0,(k-p_1)^2}^{(2)}$ and $I_{1,(k-p_1)^2}^{(2)}$ part to produce the tennis court, plus extra integrals with lower number of propagators.

Explicitly, performing the cut on the amplitude gives (after integrating over the Grassmann variables of the cut legs to account for all the particles which can run in the cut loop)
\begin{align}\label{ampsl}
\left. M^{(3)}_4(\bar{1},2,\bar{3},4)\right|_{s-cut} = i\, \frac{s\, \Tr(k\,p_1\,p_4)}{(k-p_1)^2(k+p_4)^2}\, 
\ M^{(2)}_4 (p_1,p_2,-k+p_{12},-k)
\end{align}
The final term $M^{(2)}_4$ contains the two-loop integrals in the $s$- and $(k-p_1)^2$-channel. We analyse the $s$ part first, which gives 
\begin{equation}\label{eq:cut-s}
M^{(3)}_4\, \Big|_{\text{$s$-cut, $s$-piece}} = i\, \frac{s\, \Tr(k\,p_1\,p_4)}{(k-p_1)^2(k+p_4)^2}\, \left[ I^{(2)}_{0,s} + I^{(3)}_{1,s} \right]
\end{equation}
The topology of the integrals indicates that the ladder $I^{(3)}_{lad,s}$ should contribute to this part of the cut, identifying $k=l_3$ with reference to the internal momentum labels of figure \ref{fig:lad-tc}.
Dual conformal invariance suggests that the ladder integral should appear with the numerator in \eqref{eq:ladders}. Guided by this expectation, we act with Schouten identities on \eqref{eq:cut-s} in order to reproduce this numerator.
In particular we apply the identity (which holds for generic $q$, $l_3$ is the on-shell cut momentum)
\begin{align}\label{eq:identity}
& \frac{\varepsilon\left( l_3,p_1,p_4 \right)}{(l_3-p_1)^2} \left[ s\, \varepsilon \left( q, p_1, -l_3 \right) + q^2 \varepsilon\left( p_1, p_2, -l_3 \right) \right]
=\\&
=\frac{\varepsilon\left( l_3,p_1,p_4 \right)}{t} \left[ s\, \varepsilon \left( q, p_1, p_4 \right) + q^2 \varepsilon\left( p_1, p_2, p_4 \right) \right] + s\, (l_3+p_4)^2 (q-p_1)^2 \nonumber
\end{align}
twice, onto the numerators $N$ coming from $I^{(2)}_0$. 
Indeed, after some algebra, this procedure gives rise to the desired term coming from the ladder, along with other extra pieces. These can be further manipulated in such a way that they are recognized to emerge from the cuts of the integrals $I^{(3)}_1$ and $I^{(3)}_{2}$, which consistently are also dual conformal invariant.
In particular we obtain
\begin{equation}\label{eq:scutcpiece}
-i\,  M^{(3)}_4\, \Big|_{\text{$s$-cut, $s$-piece}} = I^{(3)}_{lad,s} + I^{(3)}_{1;1} + I^{(3)}_{1;3} - I^{(3)}_{2,s} \, \Big|_{\text{$s$-cut, $s$-piece}}
\end{equation}
Acting similarly on the part of \eqref{ampsl} containing two-loop integrals in the $(k-p_1)^2$-channel we find
\begin{equation}\label{eq:cutt}
-i\, M^{(3)}_4\, \Big|_{\text{$s$-cut, $(k-p_1)^2$-piece}} = I^{(3)}_{tc,3} + I^{(3)}_{3;3} + I^{(3)}_{4;3,1} + I^{(3)}_{4;3,2} - I^{(3)}_{5;3} \, \Big|_{\text{$s$-cut, $(k-p_1)^2$-piece}}
\end{equation}
The procedure we applied, repeated on all the channels, fixes completely the integrals appearing in the amplitude, possessing two-particle cuts.

At this level we can conclude that the part of the amplitude with such cuts reads
\begin{equation}\label{eq:ampint}
-i\, M^{(3)}_4 = \frac12\, I^{(3)}_{lad,s} + I^{(3)}_{tc,1} + I^{(3)}_{1;1} - \frac12\, I^{(3)}_{2,s} + I^{(3)}_{3;3} + I^{(3)}_{4;1,4} + I^{(3)}_{4;1,3} - I^{(3)}_{5;1} + {\rm cyclic}
\end{equation}
Extra factors take care of the symmetries of integrals under cyclic permutations.
Of course this does not tell us anything about integrals without two-particle cuts, which, as remarked in the above section, are a considerable number.

We could try to infer their topologies (or exclude their presence) by performing other cuts. For instance a four-particle cut isolating two tree level six-point amplitudes or a three-particle cut separating a four-point one-loop and a tree level six-point amplitudes would provide strong consistency checks.
However they involve several contributions at the level of the integrals and the nasty six-point amplitudes on the amplitude side.  
We shall not perform such intricate cuts. Rather we pragmatically conjecture that the ladder and tennis court integrals provide the infrared divergent part of the amplitude and the functions with highest degree of transcendentality and see if we obtain a consistent answer for the amplitude. 

\paragraph{Ladder integrals}

We close this section with a remark on multi-loop ladder integrals. As discussed in the previous section, we can provide an iterative construction for the numerator of dual conformal invariant ladders.
It remains to be checked whether this is the correct numerator as appearing in $L$-loop amplitudes.
In order to check this we can generalize the two-particle cut construction above iteratively.

We perform a cut separating the four-point $L$-loop amplitude $M^{(L)}_4$ into a tree and a $(L-1)$-loop amplitudes.
Then we focus on the $s$-channel part of $M^{(L-1)}_4$ in the cut, as in \eqref{eq:scutcpiece}, which we expect to give rise to the ladder topology. We can straightforwardly repeat the steps above, and apply \eqref{eq:identity} $L-1$ times on the numerator. This produces a term coming from the cut of the numerator of \eqref{eq:Lladder} and extra pieces.
We do not have a general prediction for such additional terms, though we can imagine they could correspond to similar topologies as for the two- and three-loop cases.
In order to achieve a deeper insight into these integrals, we carry out the complete computation of the $s$-part of the cut for the four-loop amplitude. 
First we act three times on the numerator with \eqref{eq:identity} to reproduce the cut of a dual conformal invariant four-loop ladder of the form \eqref{eq:Lladder}.
We manipulate the remainders in such a way that we can identify their topology and numerators as follows
\begin{align}
M^{(4)}_4\, \Big|_{\text{ladder part}} \sim & \left(N_1+N_2\right)\, \raisebox{-5mm}{\includegraphics[width=0.2\textwidth]{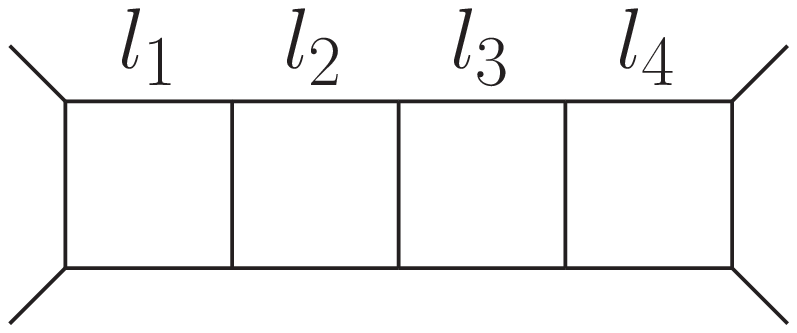}} + N_3\, \raisebox{-5mm}{\includegraphics[width=0.2\textwidth]{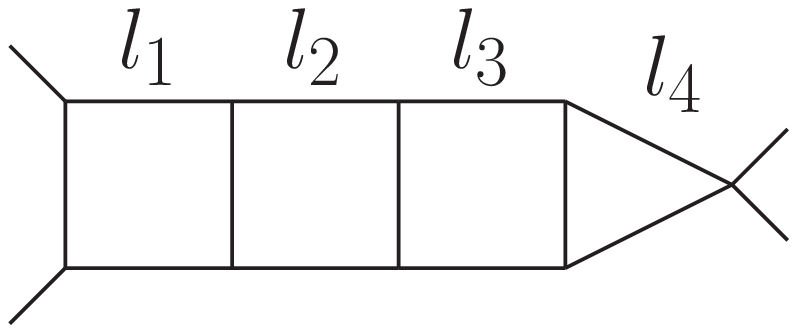}} + \nonumber\\&
+ N_4\, \raisebox{-5mm}{\includegraphics[width=0.2\textwidth]{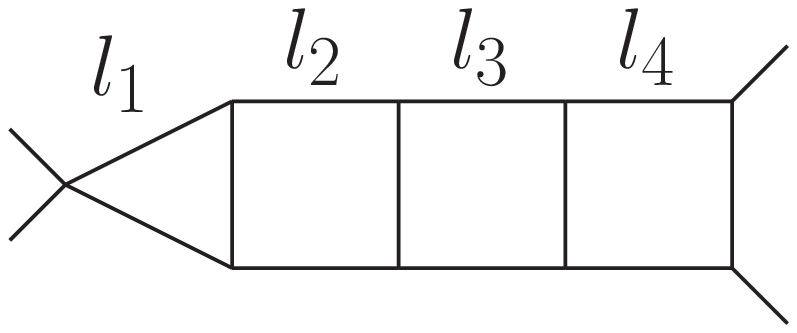}}
+ \left(N_5+N_6\right)\, \raisebox{-2.5mm}{\includegraphics[width=0.2\textwidth]{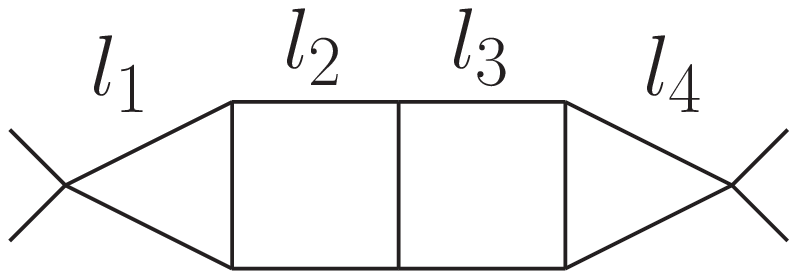}}
\end{align}
where the corresponding numerators have been defined as
\begin{align}
N_1 &= \frac{N(l_1)\, N(l_2)\, N(l_3)\, N(l_4)}{t^3} \label{eq:numladL}\\
N_2 &= \frac{s^2}{t^2}\, N(l_1)\, N(l_4)\, (l_2-p_1)^2\, (l_3+p_4)^2 \label{eq:numlad2}\\
N_3 &= \frac{s^2}{t^2}\, N(l_1) \left[ N(l_2)\, (l_3-p_1)^2 - N(l_3)\, (l_2-p_1)^2  \right] \\
N_4 &= \frac{s^2}{t^2}\, N(l_4) \left[ N(l_3)\, (l_2+p_4)^2 - N(l_2)\, (l_3+p_4)^2  \right] \\
N_5 &= \frac{s^2}{t^2}\, N(l_2) \, N(l_3) \label{eq:num4}\\
N_6 &= \frac{s^4}{t}\, (l_3-p_1)^2\, (l_2+p_4)^2 \label{eq:num5}
\end{align}
We observe in particular the appearance \eqref{eq:numladL} of the four-loop ladder \eqref{eq:Lladder}, along with another four-loop ladder topology with a simpler numerator \eqref{eq:numlad2}.
A similar combination seems to pop up for the last topology with numerators \eqref{eq:num4} and \eqref{eq:num5}. 
Consistently all these numerators make the corresponding integrals dual conformal invariant.
It would be interesting to check whether such a combination of integrals possesses uniform transcendentality as occurs at two loops.
Despite the numerators of the four-loop case turn out to be a bit more composite than at lower order, once the ladder is constructed it looks like the remaining pieces can be fixed in terms of integrals where only the extremal rungs are removed (of course other topologies where other propagators are eliminated are present if one expands numerators).
It would be interesting to check whether such a pattern persists at higher loop order.

\section{Regularization}\label{sec:regularization}

The loop integrals we have found in the previous section are in general infrared divergent and have to be regularized.
There are basically two strategies which have been proposed for amplitudes in ABJM: dimensional and Higgs mechanism regularization.
Dimensional regularization is subtle because of the ubiquitous presence of $\varepsilon$ tensors which are in principle defined in integral dimension only. Moreover, already at one loop, integrals exhibit power-like divergences, and it is not immediate that dimensional regularization could treat them properly.
Nevertheless it has proved successful in the perturbative computation of the known loop amplitudes in ABJM, as well as for various kinds of Wilson loops.
In particular it is crucial for unitarity to work properly, since the one-loop amplitude evaluated in dimensional regularization is subleading in $\epsilon$, rather than exactly vanishing, which would pose serious problems for constructing amplitudes via cuts \cite{Agarwal:2008pu}.
In particular the determination of integrands which are consistent at any order in the $\epsilon$ expansion seems to be pivotal for the application of unitarity in the computation of higher loop amplitudes.

In the following we solve the integrals identified in the previous section in dimensional regularization in $d=3-2\epsilon$ dimensions.
As anticipated, care has to be taken when dealing with $\varepsilon$ tensors. In order to fully exploit their antisymmetry property, we directly perform tensor integrals. This has the advantage that several contributions can be discarded thanks to the antisymmetry of $\varepsilon$ tensors, already at the level of Feynman parametrization. The disadvantage consists in possible errors introduced by continuing the dimension of loop momenta but keeping the $\varepsilon$ tensors in three dimensions.

The safest way to handle them would be to couple them in pairs, which can be transformed into scalar products. Even if an odd number of them is present in our numerators, we can always multiply and divide by $\varepsilon\left( p_1, p_2, p_4 \right)$ and use this to write scalar products involving loop momenta. These can be eventually translated into inverse propagators and the resulting integrals can be solved directly or reduced to master integrals by integration by parts identities.
For the two-loop amplitude this was carried out with success by solving the scalar integrals directly \cite{CH} and by reduction to master integrals \cite{Bianchi:2013pfa}.
Remarkably this result coincides with that obtained by a Feynman diagram approach \cite{BLMPS1} which produces completely different integrals without explicit $\varepsilon$ tensors in their numerators.
Even though the route outlined above is doable in principle, for the three-loop ladder and tennis court the expansion of the $\varepsilon$ tensors into inverse propagators produces a rather large amount of contributions with up to four powers of them in the numerators.
We tried to manage them by reduction to master integrals using the automatized \texttt{FIRE} routine \cite{Smirnov:2008iw}, but this failed to give an answer in a reasonable time (at least on our computer, which is soon made run out of memory).

At two loops we checked that the result of the amplitude when $\varepsilon$ tensors are previously transformed to inverse propagators coincides with performing tensor integrals and applying DRED rules for contracting metric tensors arising in the computation.
This is pleasing as DRED scheme is somehow the standard choice in supersymmetric theories and was proved to preserve gauge invariance for Chern-Simons-matter theories up to two loops \cite{Chen:1992ee}.
Remarkably this prescription also allows to derive a uniformly transcendental expression for the expectation value of a light-like Wilson loop with four cusps, that exactly agrees with the two-loop amplitude \cite{Bianchi:2013pva}. Also, when this scheme is used for the perturbative computation of supersymmetric circular Wilson loops in ABJM it provides agreement with exact results from localization \cite{Bianchi:2013zda,Bianchi:2013rma,Griguolo:2013sma}.
In all such cases application of DRED rules in the presence of $\varepsilon$ tensors can be traded with an $\epsilon$ dependent correcting factor. 
When this multiplies some pole $\epsilon^{-a}$ it then affects higher order terms ${\cal O}(\epsilon^{-a+1})$ in the $\epsilon$ expansion, by a contribution with lower degree of transcendentality.

For the scattering amplitude we do not expect three-loop integrals to exhibit severe infrared divergences, as the following argument implies. 
In fact, part of this work is devoted to ascertain whether infrared divergences exponentiate in ABJM in a similar manner to four dimensions.
Assuming this is the case, we do not expect genuine three-loop singularities, since the cusp anomalous dimension does not receive corrections at odd loops. In other words $\log {\rm Div}^{(3)} = 0$.
This means that potential infrared divergences should only come from the exponentiation of lower order corrections. In particular at three loops such contributions come from the product of the one- and two-loop amplitudes. Since the former is order $\epsilon$ and the latter possesses leading $\epsilon^{-2}$ poles, we expect simple poles at most at three loops.
This means that potential errors introduced by a sloppy treatment of $\varepsilon$ tensors would produce lower transcendentality terms in the finite part, which should be easily identified and corrected.
We shall follow this rather pragmatic strategy in the next section and verify that it gives a reasonable answer. Of course an alternative analysis would be helpful to check our results.

As an alternative scheme, regularizing by moving the theory away from the origin of moduli space was proposed in four dimensions in \cite{Alday:2009zm}. It has the theoretical appeal of preserving dual conformal invariance and the technical advantage of working in integral dimension.
It was first formulated for the ABJM theory in \cite{CaronHuot:2012hr} and applied for solving the relevant integrals for the four- and six-point two-loop amplitudes.
In that article it was shown how to implement it directly within the five-dimensional formalism, which offers several practical benefits in the solution of integrals.
At two loops it was shown that the two regularizations provide rather different results for the individual integrals, but the same final answer for the amplitude \cite{CaronHuot:2012hr}.

Regulating the integrals on the Higgs branch also provides a cleaner interpretation of the $\varepsilon$ in the numerators of integrals as dimension is fixed and not analytically continued.
Employing this method directly in five dimensions allows to work with five-dimensional $\varepsilon$ tensors \cite{CaronHuot:2012hr}. The Higgs mechanism regularization then amounts (up to logarithmic accuracy) to a shift in one of the extra-dimensional coordinates of the external points, by a mass parameter $\mu$.  
Whenever an odd number $n$ of $\varepsilon$ tensors is present in the numerator, it is always possible to express $n-1$ of them in terms of scalar products. Hence, with four external points, the final integral will always contain a numerator of the form $\varepsilon(X,X_1,X_2,X_3,X_4)$. Any integral depending on $X_1$ \dots $X_4$ only will inevitably evaluate to 0. 

As we recalled above, three-loop dual conformal invariance forces an odd number of $\varepsilon$ tensors in the numerators of integrals. Hence in massive regularization the three-loop four-point amplitude has to vanish.
Nevertheless the integrand is still non-trivial, and, as at one loop, it is crucial for unitarity at higher loop order.

\section{The amplitude in dimensional regularization}

We evaluate the integrals contributing to the two-particle cut within dimensional regularization.
We use Mellin-Barnes representation and perform it directly on the tensor integrals.
As explained in the previous section care has to be taken due to the $\varepsilon$ tensors in the numerators, which can introduce incorrect terms in the finite part of divergent integrals, if treated improperly.
On the other hand, their presence reduces significantly the number of integrals since they make some terms vanish thanks to their antisymmetry.

For instance the ladder integral can be handled as follows. First we can rewrite some factors $N(l)$ in an equivalent and more convenient form using identities \eqref{eq:identitynum} such as
\begin{equation}
s\, \varepsilon (l_1,p_1,p_4) + l_1^2\, \varepsilon (p_1,p_2,p_4) =
t\, \varepsilon (l_7,p_1,p_2) + l_7^2\, \varepsilon (p_1,p_2,p_4) 
\end{equation}
Then we can split the numerators in \eqref{eq:ladders} to give rise to four different kinds of integrals, e.g.\ corresponding to the pieces
\begin{align}
N(l_1)\, N(l_2)\, N(l_3) =& \underbrace{t^2\, \varepsilon (l_7,p_1,p_2)\, N(l_2)\, \varepsilon (l_{10},p_3,p_4)}_{a)}
+ \underbrace{t\, l_7^2\, \varepsilon (p_1,p_2,p_4)\, N(l_2)\, \varepsilon (l_{10},p_3,p_4)}_{b)}
\nonumber\\&
+ \underbrace{t\, \varepsilon (l_7,p_1,p_2)\, N(l_2)\, l_{10}^2\, \varepsilon (p_1,p_2,p_4)}_{c)}
+ \underbrace{l_7^2\, l_{10}^2\, \varepsilon^2 (p_1,p_2,p_4)\, N(l_2)}_{d)}
\end{align}
We observe that the last numerator $d)$ yields a contribution which is proportional to $I^{(3)}_{2}$, therefore we defer its evaluation.
Moreover the second and third integrals $b)$ and $c)$ are equal and each of them can only give rise to two contributions, after making a Feynman parametrization and exploiting the antisymmetry of $\varepsilon$ tensors.
The first integral $a)$ is the most complicated. Nevertheless the three-tensor part from the first piece of $N(l_2)$ containing a vector $l_2$ only produces four non-vanishing terms, whereas the two-tensor piece containing $l_2^2$ only gives two.  
Each of these contributions involves up to seven Mellin-Barnes integrations.
We perform analytic continuation and $\epsilon$ expansion using the package \texttt{MB} \cite{Czakon:2005rk}. The output of the continuation routine is commonly a rather large list of multiple Mellin-Barnes integrals. These are four-fold at most, but can always be reduced to one-fold integrals by means of repeated application of Barnes lemmas, by hand.
In some cases we find faster and more convenient to use the alternative routine \texttt{MBresolve} \cite{Smirnov:2009up}.

Remarkably we can combine all the Mellin-Barnes representations of the different pieces of the ladder, into quite a compact integral. Only a few contributions generate infrared divergences. Inspecting the corresponding tensor structure we see that they all come from pieces of the form
\begin{equation}
\tilde\eta_{\mu_1\mu_2}\, \varepsilon^{\mu_1\nu_1\rho_1}\, \varepsilon^{\mu_2\nu_2\rho_2}\, p_{i\, \nu_1}\, p_{j\, \rho_1}\, p_{k\, \nu_2}\, p_{l\, \rho_2}
\end{equation}
leading to a contraction between two $\varepsilon$ tensors. Since the metric comes from a tensor integral, we understand it as $(3-2\epsilon)$-dimensional, which we denote by $\tilde \eta$. Then we express the $\varepsilon$ tensors in terms of three-dimensional metrics and we use the DRED rule 
\begin{equation}
\tilde \eta^{\mu\nu}\, \eta_{\mu\nu} = 3-2\epsilon
\end{equation}
to contract them. 
Effectively this is equivalent to performing all algebra in three-dimensions, provided a correction factor $(1-2\epsilon)$ is introduced.
Taking into account this subtlety we can write the complete Mellin-Barnes representation of the ladder integral as
\begin{align}\label{eq:ladderMB}
I^{(3)}_{lad,s} & = 
-\frac{i\, \pi}{(4\pi)^{9/2-3\epsilon}}\, \varepsilon\left(p_1,p_2,p_4\right)\, \int_{-i\infty}^{+i\infty} \frac{dz}{2\pi i}\, (-s)^{z+1} (-t)^{-z-\frac{5}{2}}\, \Gamma^2 \left(-z-3/2\right) \Gamma (-z-1) 
\nonumber\\& ~~
\Gamma^2 (z+2) \Gamma \left(z+5/2\right)
\bigg[\frac{2}{\epsilon } - 8 \gamma_E + 8 \log 2 - \log (-s) - 5 \log (-t) + 
\nonumber\\& ~~~~~~
- 2 \psi\left(-z-3/2\right) - 3 \psi(-z-1) - 2 \psi(z+2) + 5 \psi\left(z+5/2\right)\bigg] + {\cal O}(\epsilon)
\end{align}
The other integrals emerging from this part of the two-particle cut, $I^{(3)}_1$ and $I^{(3)}_2$, turn out to be subleading in the dimensional regularization parameter.

For the tennis court the evaluation turns out to be considerably much more involved.
Even exploiting the antisymmetry of $\varepsilon$ tensors we are still left with a plethora of integrals.
Summing all contributions and after many simplifications we arrive again at a rather simple Mellin-Barnes integral. This already appears like a miracle, considering the huge expression one has to start with. Explicitly we find
\begin{align}\label{eq:tcMB}
I_{tc,s} &= -\frac{i\, \pi}{(4\pi)^{9/2-3\epsilon}}\, \varepsilon\left( p_1,p_2,p_4 \right)\, 
\frac{6 \pi ^{3/2}}{(-s)^{\frac12} t} + 
\nonumber\\& ~~~
+ \frac{i\, \pi}{(4\pi)^{9/2-3\epsilon}}\, \varepsilon\left( p_1,p_2,p_4 \right)\, \int_{-i\infty}^{+i\infty} \frac{dz}{2\pi i}\, (-s)^{z+1} (-t)^{-z-5/2}\, \Gamma \left(-z-3/2\right)^2 \Gamma (-z-1)
\nonumber\\& ~~~
\Gamma (z+2)^2 \Gamma \left(z+5/2\right) 
\bigg[-\frac{1}{\epsilon } + 4 \gamma_E + 14 + 2 \log (-s) + \log (-t) + 
\nonumber\\& ~~~~~~~
- 2 \psi\left(-z-3/2\right) + 4 \psi(z+2) - \psi \left(z+5/2\right)
\bigg] + {\cal O}(\epsilon)
\end{align}
We can observe the presence of terms with lower transcendentality.
Still we have to combine this with the other integrals originating from the two-particle cut, according to \eqref{eq:ampint}.
Their evaluation is simpler and gives respectively
\begin{align}
I^{(3)}_{3;1} &= -\frac{16\, i\, \pi}{(4\pi)^{9/2-3\epsilon}}\, \int_{-i\infty}^{+i\infty} \frac{dz}{2\pi i}\, (-s)^{z+1} (-t)^{-z-\frac{5}{2}} \Gamma^2 \left(-z-3/2\right) \Gamma (-z-1) 
\nonumber\\& ~~~~~~~~
\Gamma^2 (z+2) \Gamma \left(z+5/2\right) + {\cal O}(\epsilon)
\nonumber\\ 
I^{(3)}_{4;1,4} &= I^{(3)}_{5;1} + {\cal O}(\epsilon) = \frac{6\, i\, \pi^{\frac52}}{(4\pi)^{9/2-3\epsilon}} \, \frac{\varepsilon\left(p_1,p_2,p_4\right)}{(-s)^{\frac12}\,t} + {\cal O}(\epsilon)
\end{align}
Considering the combination dictated by the cut condition \eqref{eq:cutt}, we see that part of the terms of lower transcendentality cancels out.
Still we are left with a lower transcendentality piece in the Mellin-Barnes integral.
Although we do not have a clean argument to justify it as for the ladder integral, we see that an extra factor $(1-2\epsilon)$ would provide a perfect cancellation of this last piece of lower transcendentality. We therefore postulate this to be the correct prescription and write
\begin{align}\label{eq:tcMBfin}
& I^{(3)}_{tc,1} + I^{(3)}_{3;1} + I^{(3)}_{4;1,3} + I^{(3)}_{4;1,4} - I^{(3)}_{5;1} = -\frac{i\, \pi}{(4\pi)^{9/2-3\epsilon}}\, \varepsilon\left( p_1,p_2,p_4 \right)\,
\int_{-i\infty}^{+i\infty} \frac{dz}{2\pi i}\, (-s)^{z+1} (-t)^{-z-\frac52} 
\nonumber\\& ~~
\Gamma \left(-z-3/2\right)^2 \Gamma (-z-1) \Gamma (z+2)^2 \Gamma \left(z+5/2\right) 
\bigg[\frac{1}{\epsilon } - 4 \gamma_E - 2 \log (-s) - \log (-t) +
\nonumber\\& ~~~~~~~~~~
+ 2 \psi\left(-z-3/2\right) - 4 \psi(z+2) + \psi \left(z+5/2\right)
\bigg] + {\cal O}(\epsilon)
\end{align}
It would be interesting to check this result against a different solution of the integrals, getting rid of $\varepsilon$ tensors from the onset.

We finally sum up all contribution \eqref{eq:ampint} to obtain a Mellin-Barnes integral representation for the amplitude, which reads
\begin{align}\label{eq:ampMB}
M^{(3)}_4 &= \frac{1}{(4\pi)^{1/2-3\epsilon}}\, \varepsilon\left( p_1,p_2,p_4 \right)\,
\int_{-i\infty}^{+i\infty} \frac{dz}{2\pi i}\, (-s)^{z+1} (-t)^{-z-5/2} \Gamma \left(-z-3/2\right)^2 \Gamma (-z-1) 
\nonumber\\& ~~~~
\Gamma (z+2)^2 \Gamma \left(z+5/2\right) \bigg[
\frac{2}{\epsilon } - 8 \gamma_E + 4 \log 2 - 3 \log (-s) - 3 \log (-t) + 
\nonumber\\& ~~~~~~~~~~~~
- 2 \psi\left(-z-3/2\right) + \psi\left(-z-1\right) - 2 \psi(z+2) + \psi \left(z+5/2\right)
\bigg] + {\cal O}(\epsilon)
\end{align}

\section{The $\epsilon$ expansion of the one-loop amplitude}

The one-loop contributions to the ABJM four-point amplitude is known to be subleading in $\epsilon$ when evaluated in dimensional regularization.
Here we provide an explicit expression for its expansion in the dimensional regularization parameter $\epsilon$ up to order $\epsilon^2$.
Since the two-loop amplitude has leading $\epsilon^2$ poles, this is the order required to evaluate completely the logarithm of the amplitude $\log {\cal M}$ up to finite order at three loops.

By contraction with $\varepsilon\left( p_1, p_2, p_4\right)$ we can decompose the one-loop box function into a scalar box and two triangles. Both can be given an all-order expression in $\epsilon$. For the box this can be done borrowing four-dimensional results and performing the shift $\epsilon \rightarrow \epsilon + 1/2$. The result is expressed in terms of $_3 F_2$ hypergeometric functions, which can be in principle expanded in $\epsilon$. Such an expansion, however, is not straightforward. As an alternative path we can write down the Mellin-Barnes representation of the box-function and expand it up to order ${\cal O}(\epsilon^2)$
\begin{align}\label{eq:1loopMB}
M_4^{(1)} & = \epsilon^2\, \frac{2\, e^{-2\gamma_E \epsilon}}{(4\pi)^{1/2-\epsilon}}\, \varepsilon\left(p_1,p_2,p_4\right)\,
\int_{-i\infty}^{+i\infty} \frac{dz}{2\pi i}\, (-s)^{z+1}\, (-t)^{-z-\frac{5}{2}}\,\Gamma^2 \left(-z-3/2\right) \Gamma \left(-1-z\right) 
\nonumber\\&
~~~~~~~~ \Gamma^2 \left(z+2\right) \Gamma \left(z+5/2\right)
\bigg[ -\frac{(-s)^{-\epsilon} + (-t)^{-\epsilon}}{\epsilon} 
+2 \psi \left(-z-3/2\right)-\psi\left(-z-1\right)+
\nonumber\\&
~~~~~~~~~~~~~~~~~~~~~~~ + 2 \psi\left(z+2\right)-\psi\left(z+5/2\right)
\bigg] + {\cal O}(\epsilon)
\end{align}
in a form which is manifestly invariant under $s\leftrightarrow t$, which is equivalent to the shift of the integration variable by $z\rightarrow -z-7/2$.
For the order $\epsilon$ term we can directly pick residues up and sum the corresponding series. In order to explicitly evaluate the Mellin-Barnes integral at order $\epsilon^2$ we went back to the Feynman parametrized form of the box integral and expanded it to such order. The resulting integral can be solved in terms of classical polylogarithms. 
We did not find closed expressions in terms of polylogarithms for the individual series underlying the $\epsilon^2$ expansion of \eqref{eq:1loopMB}.

Before presenting the result, we introduce some convenient variables $x\equiv \sqrt{\frac{s+t}{s}}$, $y\equiv \sqrt{\frac{s+t}{t}}$ and remark that we shall focus on the Euclidean region $s<0$, $t<0$.
The overall factor $\varepsilon\left(p_1,p_2,p_4\right)$ can also be expressed in terms of Mandelstam variables as $\pm \frac12 \sqrt{-s\, t\, (s+t)}$, where the sign depends on the kinematics. We fix the minus sign and use the above identification to get finally rid of such factors.
Finally the ${\cal O}(\epsilon^2)$ order expansion of the one-loop box function reads 
\begin{align}
M_4^{(1)} & = 2\, \pi\, \left(16\, e^{-\gamma_E}\, \pi\right)^{\epsilon}\, \bigg\{
\epsilon\,  \left[(-s)^{-\epsilon } \text{ArcTanh} \,y^{-1} + (-t)^{-\epsilon } \text{ArcTanh} \,x^{-1} \right] + \label{eq:1loopdiv}
\\&~~~~~~~~
-\epsilon ^2 \bigg[
\text{Li}_2(1-x) + \text{Li}_2(-x) + \text{Li}_2(1-y) + \text{Li}_2(-y) 
\nonumber\\&~~~~~~~~~~~~
+ \log (x-1)\, \log \, x  + \log (y-1)\, \log \, y - \frac{\pi ^2}{3}\bigg]
\bigg\} + {\cal O}\left(\epsilon^3\right) \label{eq:1loopfin}
\end{align}
We observe that the functions appearing in this expansion exhibit uniform degree of transcendentality.

\section{Conclusions: exponentiation and polylogarithms}

In light of the results of the previous section we comment on the properties of our conjectural expression for the three-loop four-point amplitude \eqref{eq:ampMB}.
\begin{itemize}
\item
We can easily ascertain from the Mellin-Barnes representations \eqref{eq:ampMB}, \eqref{eq:1loopMB} and the infrared divergences of the two-loop amplitude \eqref{eq:2loopamp} that
\begin{equation}\label{eq:expdiv}
{\cal M}^{(3)}_4\, \Big|_{div}\, =\, {\cal M}^{(1)}_4\, \times\, {\cal M}^{(2)}_4\, \Big|_{div} 
\end{equation} 
In particular we can verify that infrared divergences exponentiate to this perturbative order.
This is because there is no three-loop contribution to the cusp anomalous dimension and therefore we expect all three-loop infrared divergences to be caused by lower order corrections. These should therefore cancel out when considering $\log\, {\cal M}$, as \eqref{eq:expdiv} shows it is the case.
Moreover all the pieces in the finite part of the amplitude (see \eqref{eq:result}) of the form $\log(-s)$ can be finally packaged into dipole variables $(-s)^{l\epsilon}$. These are the same natural objects appearing in exponentiation of infrared divergences in four dimensions. 

On the one hand this exponentiation can sound rather expected, by analogy with what occurs in four dimensions. On the other hand any previous computation of amplitudes in ABJM was confined to the lowest perturbative order where divergences turn up, and consequently does not provide sufficient hints in favour of such an exponential behaviour.
If we believe that infrared divergences have to exponentiate, for instance following the arguments of \cite{Huang:2014xza}, this suggests that our result \eqref{eq:ampMB} could in fact give the whole three-loop amplitude (at least up to subleading order ${\cal O}(\epsilon)$ contributions), despite it was derived only from integrals possessing a two-particle cut. 
Indeed the fact that all physical infrared divergences are captured by our ansatz excludes other divergent contributions to the amplitude. Since functions with highest degree of transcendentality usually appear within divergent integrals, we suspect that at least the highest transcendentality part of the amplitude is covered by our result. Of course some extra divergent integrals could conspire in such a way that their divergent parts cancel out, leaving some additional contributions. We can not exclude their presence at this stage, but the consistency of the results and analogous lower loop computations give strong evidence this is probably not the case.

A remark is in order: even if our conjecture for the three-loop amplitude was correct, this would not necessarily mean that its integrand coincides with that found in \eqref{eq:ampint}. With our two-particle cut analysis and assuming that the amplitude possesses dual conformal invariance, we can only assert to have completely determined the part of the integrand which is sensitive to such cuts.
This does not exclude the possibility that finite dual conformal invariant integrals without such a cut can appear in the integrand.
We have seen that already in our expression \eqref{eq:ampint} two integrals are subleading in $\epsilon$. Henceforth any other subleading integral or combination thereof could be present at the level of the integrand without modifying \eqref{eq:ampMB} to finite order.
Such integrands would however play an important role if the three-loop amplitude is used as an input for a higher order unitarity-based computation. 

\item
Including finite terms, we obtain the remarkable result that
\begin{equation}\label{eq:exp}
{\cal M}^{(3)}_4\, =\, {\cal M}^{(1)}_4\, \times {\cal M}^{(2)}_4 + {\cal O}(\epsilon)
\end{equation}
This result is compatible with an exponentiation ansatz for the four-point amplitude in the spirit of \cite{BDS}.
Indeed to the order we are working we can write
\begin{equation}
\log\, {\cal M}_4 = 
 {\cal M}^{(2)}_4 + {\cal O}(\epsilon) + {\cal O}\left(\lambda^4 \right)
\end{equation}
As proposed in \cite{CH,BLMPS1,Bianchi:2011aa} one can extend this formula to higher order and conjecture a BDS ansatz for the four-point amplitude
\begin{equation}\label{eq:ansatz}
\log\, {\cal M}_4 = \sum_{l=1}^{\infty}\, f_{CS}^{(2l)}(\epsilon)\, {\cal M}^{(2)}_4(l\epsilon) + C^{(l)} + {\cal O}(\epsilon)
\end{equation}
where $f_{CS}$ is a function whose $\epsilon$-independent part gives the coefficients of the scaling function of ABJM theory \cite{GV,MOS1,MOS2,LMMSSST}, and $C^{(l)}$ are possible constant non-iterating contributions \cite{BDS}.

A striking difference with respect to the four-dimensional case is that the logarithm of the amplitude would be expressed in terms of the two-loop amplitude instead of the one-loop contribution, as for ${\cal N}=4$ SYM. We recall that the main ingredient of \eqref{eq:ansatz}, namely the two-loop amplitude in ABJM, coincides with the ${\cal N}=4$ SYM one-loop amplitude, up to terms which can be reabsorbed in a redefinition of $f$ and $C$ and subleading contributions in $\epsilon$. Henceforth \eqref{eq:ansatz} is actually strikingly similar in form to the BDS ansatz for the four-point amplitude of ${\cal N}=4$ SYM.

We stress that taking the logarithm of the amplitude hides considerable information in the ${\cal O}(\epsilon)$ terms, which are crucial for recovering higher loop results when exponentiating. In particular the whole one-loop amplitude falls into the subleading pieces and disappears from the ansatz \eqref{eq:ansatz}, although we have seen that it plays a crucial role for finding the three-loop correction from it. This observation considerably sharpens previous proposals for a BDS exponentiation \cite{Bianchi:2011aa}, where possible order $\epsilon$ terms correcting the two-loop amplitude where included in the exponential. Here we discover that such terms are precisely given by the one-loop amplitude. It would be interesting to further check this exponentiation proposal at higher order, where the second term in the perturbative expansion of the cusp anomalous dimension $f_{CS}$ appears.

In ${\cal N}=4$ SYM the BDS ansatz for the four-point amplitude can be interpreted as the result of an anomalous conformal Ward identity descending from dual conformal invariance.
The hints of an exponential structure for the four-point amplitude in ABJM provides further support in favour of this symmetry to hold at higher order in perturbation theory, despite strong coupling objections.

Since at four points conformal Ward identities are extremely constraining for amplitudes and Wilson loops, it is likely that a duality between amplitudes and Wilson loops also holds beyond two loops
\begin{equation}
\log\, \langle W_4 \rangle\, = \, \log\, {\cal M}_4 + {\cal O}(1/N)
\end{equation}
At three loops this implies that the light-like Wilson loop should vanish, since the one-loop correction is zero and hence no contributions can emerge from the exponentiation of lower order terms.
This strong prediction is a result of the conformal properties of light-like Wilson loops and the structure of cusp divergences in ABJM. It would be interesting to test this via a direct computation of the Wilson loop at three loops.

\item
Tree level amplitudes in ABJM have been given a description in terms of contour integrals in an orthogonal Grassmannian space \cite{Lee} and on-shell graphs \cite{Huang:2013owa}. 
Recently this description has been extended to loop corrections \cite{Huang:2014xza}.
This formulation entails exponentiation of infrared divergences in a similar way as in four dimensions. We have verified that this is the case for our conjecture on the three-loop amplitude.
Another remarkable feature of the Grassmannian integral formulation is that the four-point amplitude can be expressed in a $d\log$ form which implies that its $l$-loop corrections are given by functions of uniform degree of transcendentality $l$. 
Using \eqref{eq:exp} we can provide a check of this statement at three loops, as the amplitude can be clearly expressed in terms of classical polylogarithms with uniform degree of transcendentality
\begin{align}\label{eq:result}
& M_4^{(3)} = \pi\, \frac{\left(-s/\mu'^2\right)^{-3 \epsilon}+\left(-t/\mu'^2\right)^{-3 \epsilon}}{2} \bigg[
- \frac{1}{\epsilon}\, \left(\text{ArcTanh}\, x^{-1} + \text{ArcTanh}\, y^{-1} \right)
\nonumber\\& 
- \log 2\, \left(\text{ArcTanh}\, x^{-1} + \text{ArcTanh}\, y^{-1} \right)
+ \log\, \frac{x}{y}\, \left(\text{ArcTanh}\, x^{-1} - \text{ArcTanh}\, y^{-1} \right)
\nonumber\\&
+ \text{Li}_2(1-x) + \text{Li}_2(-x) + \text{Li}_2(1-y) + \text{Li}_2(-y) 
\nonumber\\&~~~~~~~~~~~~
+ \log (x-1)\, \log \, x  + \log (y-1)\, \log \, y - \frac{\pi ^2}{3}
\bigg] + {\cal O}(\epsilon)
\end{align}
where $\mu'^2 = 8 \pi e^{-\gamma_E}\mu^2$ is the same scale as defined after \eqref{eq:2loopamp} and $x$ and $y$ are the square roots of invariant ratios appearing in the expansion of the one-loop amplitude.
In fact we are not able guarantee that the three-loop ladder and tennis court can be expressed in terms of polylogarithms as well.
This would require solving explicitly their Mellin-Barnes representations \eqref{eq:ladderMB} and \eqref{eq:tcMB}.
This task looks challenging, due to the polygamma functions in the integrand and the fact that residues are taken at half-integer points. This produces rather intricate series which we shall not attempt to solve.
Still, we can observe that symmetrizing these integrals under $s\leftrightarrow t$ yields considerable simplification, as the integrands involving polygamma functions take the same combination as that appearing in \eqref{eq:1loopMB}.
Exploiting this identification and the solutions \eqref{eq:1loopdiv} and \eqref{eq:1loopfin} we can straightforwardly extract explicit results for $I^{(3)}_{lad,s} + I^{(3)}_{lad,t}$ and $I^{(3)}_{tc,1} + I^{(3)}_{tc,2}$ in terms of polylogarithms.

\end{itemize}

\section*{Acknowledgements}

We thank Lorenzo Bianchi, Gast\'on Giribet and Silvia Penati for very useful discussions. 
The work of MB has been supported by the Volkswagen-Foundation.

\vfill
\newpage

\appendix

\section{Notation and conventions}\label{app:1}

We work with the Minkowski metric $g_{\mu\nu}={\rm diag}\{1,-1,-1\}$
and the totally antisymmetric tensor $\varepsilon^{\mu\nu\rho}$, defined by
$\varepsilon_{012}=\varepsilon^{012}=1$. Spinor indices are raised and lowered as $\l_{\alpha}=\varepsilon_{\alpha \beta} \l^{\beta}$ with $\varepsilon_{12}=\varepsilon^{12}=1$.

On-shell solutions of the fermionic equations of motion are expressed in terms of $SL(2,\mathbb{R})$ commuting spinors $\l_\a$. The same quantities allow one to write on-shell momenta as 
\begin{equation}
p_{\a\b}=(\gamma^\mu)_{\a\b}\ p_\mu
\end{equation}
where the set of $2 \times 2$ gamma matrices are chosen to satisfy 
\begin{equation}
\left(\g^{\mu}\right)^{\a}_{~\g}\, \left(\g^{\nu}\right)^{\g}_{~\b} = -g^{\m\n}\, \d^{\a}_{~\b} - \e^{\m\n\rho}\, \left(\g_{\rho}\right)^{\a}_{~\b}
\end{equation}
An explicit set of matrices is $( \gamma^{\mu} )_{\a\b}= \{ \s^0, \s^1, \s^3 \}$. 

We define spinor contractions as
\begin{equation}
\langle i \, j\rangle=-\langle j \, i\rangle \equiv \lambda^{\alpha}_i\lambda_{\alpha j}=\epsilon_{\alpha\beta}
\lambda^{\alpha}_i \lambda^{\beta}_j
\end{equation}
They obey the Schouten identity
\begin{equation}\label{eq:schouten}
\braket{a b} \braket{c d} + \braket{a c} \braket{d b} + \braket{a d} \braket{b c} = 0
\end{equation}
Thus for any pair of on-shell momenta we write
\begin{equation}
p_{ij}^2 \equiv (p_i+p_j)^2 = 2 \, p_i \cdot p_j =  p_i^{\a\b} \, (p_j)_{\a\b} = \langle i \, j\rangle^2
\end{equation}
For positive energy spinors are real, whereas for negative energy they are imaginary.

Traces:
\begin{equation}
\braket{ij} \braket{ji}\ =\ -2\ p_i \cdot p_j
\end{equation}
\begin{equation}
\braket{ij} \braket{jk} \braket{ki}\ =\ \Tr (p_i\ p_j\ p_k) = 2\, \epsilon(i,j,k)
\end{equation}
\begin{align}
& \braket{ij} \braket{jk} \braket{kl} \braket{li}\ =\ \Tr (p_i\ p_j\ p_k\ p_l) = 
\nonumber\\&
2\, \left[
{p_i}.{p_j}\ {p_k}.{p_l} + {p_i}.{p_l}\ {p_j}.{p_k} - {p_i}.{p_k}\ {p_j}.{p_l}
\right]
\end{align}

For definiteness we will choose a regime where
\begin{equation}
\braket{12} = \braket{43} \quad \braket{23} = \braket{41} \quad \braket{13} = \braket{24} 
\end{equation}

We will use the four-point superamplitude
\begin{equation}
{\cal A}_4 = i\, \frac{\delta^{(3)}(P)\delta^{(6)}(Q)}{\braket{12}\braket{23}}
\end{equation}

At loop level our integrals are normalized with the measure
\begin{equation}
\int \frac{d^{3-2\epsilon}k}{(2\pi)^{3-2\epsilon}}
\end{equation}
for each loop integration.

\section{Mellin-Barnes integrals solutions}

We review here the solutions of the relevant Mellin-Barnes integrals for the expansion of the one-loop amplitude up to order $\epsilon^2$.
In particular we have
\begin{align}
& \int_{-i\infty}^{+i\infty} \frac{dz}{2\pi i}\, y^z\, \Gamma^2 \left(-z-3/2\right) \Gamma (-z-1) \Gamma^2 (z+2) \Gamma \left(z+5/2\right) = 
\nonumber\\&
= \frac{2 \pi ^{3/2}}{y \sqrt{y (y+1)}}\, \left( \text{ArcTanh}\, \sqrt{\frac{1}{y^{-1}+1}} + \text{ArcTanh}\,\sqrt{\frac{1}{y+1}}\right)
\end{align}
and
\begin{align}
& \int_{-i\infty}^{+i\infty} \frac{dz}{2\pi i}\, y^z\, \Gamma^2 \left(-z-3/2\right) \Gamma (-z-1) \Gamma^2 (z+2) \Gamma \left(z+5/2\right) \nonumber\\&
~~~~ \left[ 2 \psi\left(-z-3/2\right)-\psi(-z-1)+2 \psi(z+2)-\psi\left(z+5/2\right) \right] = 
\nonumber\\&
= \frac{2 \pi ^{3/2}}{y \sqrt{y (y+1)}}\, \left[
2\, \text{Li}_2 \left(1-\sqrt{1+y^{-1}}\right) + 2\, \text{Li}_2\left(-\sqrt{1+y^{-1}}\right) +
\right. \nonumber\\& ~~~~
+ 2\, \text{Li}_2\left(-\sqrt{y+1}\right) + 2\, \text{Li}_2\left(1-\sqrt{y+1}\right)+
\nonumber\\& ~~~~
+ 2\, \log \left(\sqrt{y^{-1}+1}-1\right) \log \, \sqrt{y^{-1}+1} + 2\, \log \, \sqrt{y+1}\, \log \left( \sqrt{y+1}-1\right)
\nonumber\\& ~~~~
-(2 \gamma_E + 4 \log 2) \left(\text{ArcTanh}\, \sqrt{\frac{1}{y^{-1}+1}} + \text{ArcTanh}\, \sqrt{\frac{1}{y+1}}\right)+ 
\nonumber\\& \left. ~~~~ 
+ \log\, y \left(\text{ArcTanh}\, \sqrt{\frac{1}{y+1}} - \text{ArcTanh}\, \sqrt{\frac{1}{y^{-1}+1}} \right) - \frac{2 \pi ^2}{3}
\right]
\end{align}

\vfill
\newpage

\bibliographystyle{JHEP}

\bibliography{biblio}

\end{document}